\documentclass[runningheads]{llncs}
\usepackage[T1]{fontenc}
\usepackage{appendix}
\usepackage{booktabs}
 \usepackage{multirow}
 \usepackage{comment}
 \usepackage{booktabs}
 \usepackage{subfigure}
 \usepackage[linesnumbered,ruled,vlined]{algorithm2e}
\usepackage{amsmath}
\usepackage{graphicx}
\usepackage{color}
\usepackage{amssymb}
\usepackage{enumitem,kantlipsum}
\usepackage{xfrac}

\begin{document}

\newcommand{\rev}[1]{{\color{blue}#1}}
\newcommand{\todo}[1]{{\color{red}#1}}
\definecolor{amethyst}{rgb}{0.54, 0.17, 0.89}
\definecolor{ao}{rgb}{0.0, 0.5, 0.0}
\newtheorem{definitionenv}{Definition}
\title{Taxonomy Inference for Tabular Data  Using Large Language Models}
%
%
\author{Zhenyu Wu\inst{1}\orcidID{0000-0003-0981-5567} \and
Jiaoyan Chen\inst{1}\orcidID{0000-0003-4643-6750} \and
Norman W. Paton \inst{1}\orcidID{0000-0003-2008-6617}}
\authorrunning{Z. Wu et al.}
%
\institute{The University of Manchester
\email{\{zhenyu.wu,jiaoyan.chen,norman.paton\}@manchester.ac.uk}}
\maketitle              
\begin{abstract}
Taxonomy inference for tabular data is a critical task of schema inference, aiming at discovering entity types (i.e., concepts) of the tables and building their hierarchy. It can play an important role in data management, data exploration, ontology learning, and many data-centric applications.
Existing schema inference systems focus more on XML, JSON or RDF data, and often rely on lexical formats and structures of the data for calculating similarities, with limited exploitation of the semantics of the text across a table. 
Motivated by recent works on taxonomy completion and construction using Large Language Models (LLMs), this paper presents two LLM-based methods for \textbf{t}axonomy inference for \textbf{t}ables: (i) EmTT which \textbf{em}beds columns by fine-tuning with contrastive learning encoder-alone LLMs like BERT and utilises clustering for hierarchy construction, and (ii) GeTT which \textbf{ge}nerates table entity types and their hierarchy by iterative prompting using a decoder-alone LLM like GPT-4.  
Extensive evaluation on three real-world datasets with six metrics covering different aspects of the output taxonomies has demonstrated that EmTT and GeTT can both produce taxonomies with strong consistency relative to the Ground Truth.

\end{abstract}
%
%
%


\section{Introduction}
\label{sec:intro}
Schema inference, which is to identify the structure, meta information and semantics of a dataset such as relationships between data fields and data types, plays a critical role in data management, ontology learning, and data-centric applications \cite{DBLP:conf/edbt/BaaziziLCGS17,DBLP:conf/edbt/BarretMU24,DBLP:conf/esws/VolkerN11}.
In particular, inferring the entity types (i.e., semantic types or concepts like \textit{School} and \textit{Hotel}) of tables in a given dataset as well as their hierarchies is one of the fundamental tasks in schema inference. It not only provides necessary semantics for other tasks like mining constraints on the relationships, but also directly supports data exploitation in quite a few scenarios such as Knowledge Graph population, table retrieval and table question answering \cite{DBLP:journals/corr/abs-2404-16130,DBLP:journals/ws/LiuCTHLM23}.

However, most of the current schema inference methods consider XML or JSON documents, or graph data composed of RDF triples  \cite{DBLP:journals/vldb/Kellou-MenouerK22,DBLP:journals/csur/PatonCW24}, while inferring type hierarchies for sets of heterogeneous tables is paid little attention.
Furthermore, early proposals mostly rely on lexical formats and the structure of the data for calculating similarities, without fully exploiting the semantics of the unstructured or semi-structured text~\cite{DBLP:journals/tlsdkcs/ChristodoulouPF15,DBLP:conf/er/Kellou-MenouerK15}.
Recently, there are some works that attempt to train neural networks, especially Transformer-based architectures, or fine-tune their pre-trained versions, for embedding tabular data for measuring similarities and conducting prediction tasks such as column type annotation and joinable table discovery (e.g., TURL \cite{DBLP:journals/pvldb/DengSL0020} and DeepJoin \cite{DBLP:journals/pvldb/Dong0NEO23}), but there is a shortage of exploration into complex tasks including hierarchy inference, which has as input a set of heterogeneous tables and a structured output, and may rely on multiple steps.

Meanwhile, quite a few recent works explore Large Language Models (LLMs) of different architectures including encoder-alone, encoder-decoder and decoder-alone for taxonomy construction and curation, utilising text. 
Most of these works focus on taxonomy completion, such as the insertion of new concepts or new subsumption relationships, by transforming the problem into machine learning classification based on the encoding of LLMs (e.g., \cite{DBLP:journals/www/ChenHGJDH23,DBLP:journals/jbi/LiuPG20,DBLP:conf/esws/ShiCDKLZWH23}). 
There are also several works that develop complex prompts, which are often iterative, for generative LLMs like the Llama \cite{DBLP:journals/corr/abs-2302-13971} and GPT series for constructing taxonomies from scratch or from a given set of concepts (e.g., \cite{DBLP:conf/kbclm/FunkHJL23,DBLP:conf/cikm/0001BTFLZ024}). 
However, none of these LLM-based works have explored taxonomy construction for a given set of heterogeneous tabular data, which involves not only the manipulation of the concepts but also learning from the raw data.  

The contributions of this work are as follows:
\begin{enumerate}
\item A proposal for an \textit{Embedding-based Method}, that clusters  column embeddings to identify top-level concepts, their attributes and concept hierarchies, in sequence, and which can be used with pretrained or fine-tuned language models.
\item A proposal for a  \textit{Generative Method} that prompts a pre-trained generative LLM such as GPT-4, DeepSeek-R1 \cite{deepseekai2025deepseekr1} or Qwen2 \cite{DBLP:journals/corr/abs-2407-10671} to infer table semantic types, and then applies an iterative prompting method named Chain-of-Layer \cite{DBLP:conf/cikm/0001BTFLZ024} for taxonomy construction.
\item An evaluation of (1) and (2), each building on several language models, on three real-world table sets with annotated ground truth hierarchies, using six metrics that consider quality of both top-level types and the overall taxonomy, with positive results.
\end{enumerate}

For conciseness, we refer to these two methods as EmTT and GeTT, respectively, where TT is short for \textbf{t}axonomy inference for \textbf{t}abular data.
{We study two methods that use different language model techniques to thoroughly explore the potential of language models for solving the problem of table hierarchy inference.}


\section{Problem Statement}

Given a set of tables $\mathcal{D}$, this study aims to: \textit{(i)} for each table $d \in \mathcal{D}$, infer entity types that the corresponding entities of the table rows all belong to; \textit{(ii)} with the entity types of all the tables, denoted as $\mathcal{T}$, build an \textit{entity type taxonomy} (taxonomy in short), denoted as \(\mathcal{H} = (\mathcal{T}, \mathcal{E})\), which is a directed acyclic graph (DAG) for representing the hierarchy of $\mathcal{T}$, with \(\mathcal{E}\) being a set of directed edges representing "is-a" (i.e., subsumption) relationships between the entity types. 
This formulation assumes that each table is associated with to a single entity type; non-entity tables can be converted into entity tables by a recent method \cite{DBLP:journals/sigmod/LiHYWC24}.

\section{Embedding-based Method}
\label{sec:EmTT}
\begin{figure}[t]
    \centering
    \includegraphics[width=0.85\linewidth]{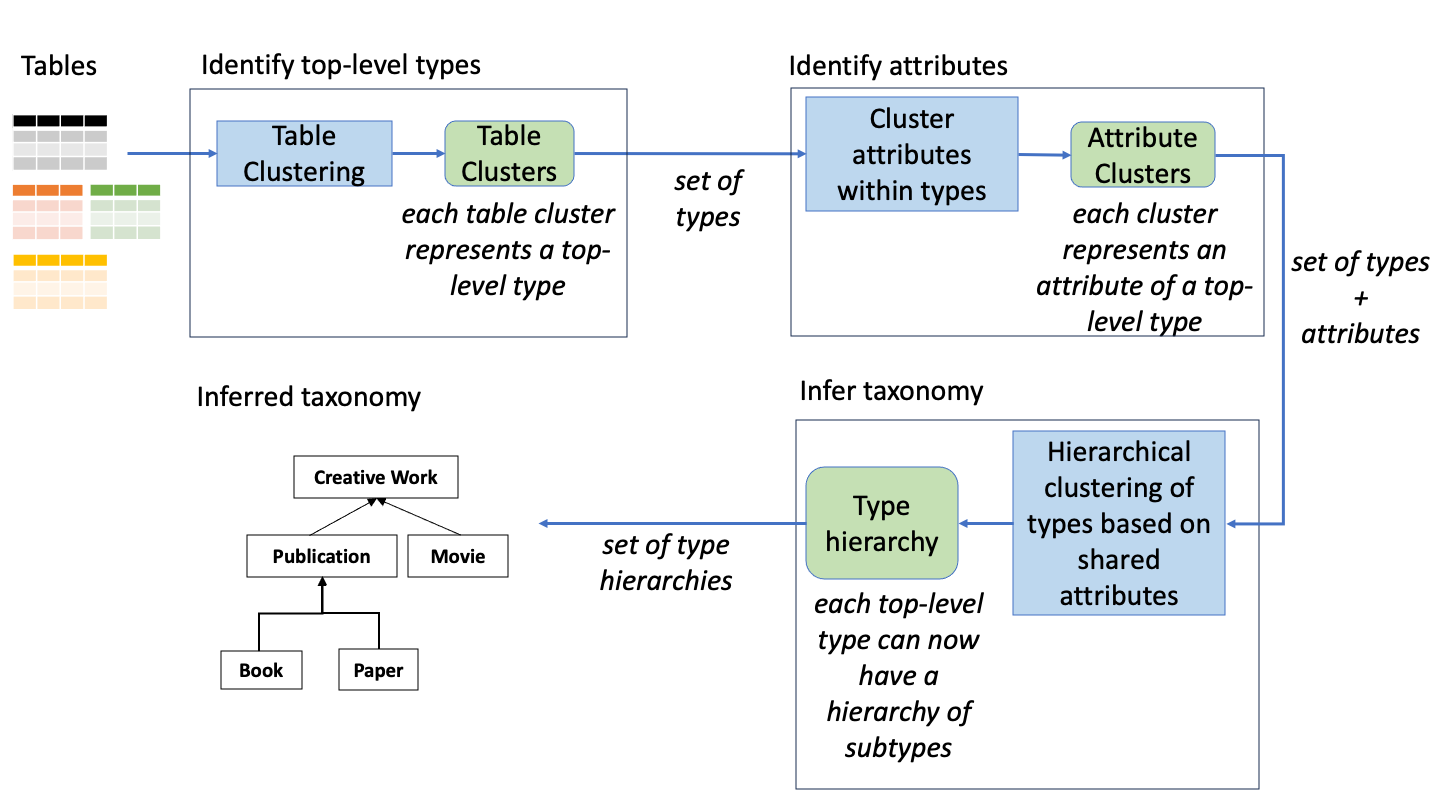}
    \vspace{-0.15cm}
    \caption{{The Framework of the Embedding-based Method EmTT}}
    \label{fig:emtt}
\end{figure}

In this section, we introduce the embedding-based method EmTT. {As illustrated in Figure \ref{fig:emtt}, it includes the following three steps: \textit{Identify top-level types}, which clusters the tables according to their column embeddings, with each cluster representing one top-level type; \textit{Identify attributes}, which clusters columns of the tables associated with each top-level type in turn, with each cluster representing an attribute of the top-level type; and \textit{Infer taxonomy}, which utilises these type attributes to group the tables of each top-level type into hierarchical sub-types based on their shared attributes.}
We now introduce each step.

\subsection{Identify Top-level Types}
\label{subsec:ttl}

Each table is composed of rows and columns (a.k.a. table attributes), with each row representing an entity.
Such flat entity tables are common in real-world data such as web pages and government data\footnote{Studies indicate that 30-50\% of spreadsheets contain non-entity tables~\cite{DBLP:journals/sigmod/LiHYWC24}. However, recent techniques have been proposed to automate the conversion of non-entity tables into entity tables~\cite{DBLP:journals/sigmod/LiHYWC24}, which broadens the applicability of methods relying on subject attributes.}. It is assumed that each table contains a column representing the \textit{entity type} (i.e., semantic type or concept) that the table is about, which is known as the \textit{subject column} or \textit{subject attribute}. {For example, in a table of \textit{companies}, the subject column could be the \textit{company name}, while the other columns provide additional properties of the company such as its head office address or turnover.
Subject columns have been used to support a variety of tasks, such as web table extraction~\cite{DBLP:journals/pvldb/VenetisHMPSWMW11} and table annotation~\cite{DBLP:conf/er/WangWWZ12}, and several methods have been proposed for their inference (e.g., ~\cite{DBLP:journals/pvldb/VenetisHMPSWMW11,DBLP:journals/semweb/Zhang17})}.

To identify top-level types, a set of clusters \( \mathcal{C} \) is created, where each \( c \in \mathcal{C} \) groups tables from $\mathcal{D}$ that share a common high-level type like \textit{Organization} or \textit{Person}.
%
The approach involves two steps: (i) identifying the subject column of each table using an existing technique proposed in \cite{DBLP:journals/semweb/Zhang17}; and (ii) clustering the tables based on the semantic similarity of the embeddings of their subject columns, where the number of clusters is chosen with the Silhouette Coefficient \cite{DBLP:journals/silhouette}. 

{In experiments, we use Agglomerative Clustering with Euclidean distance as the metric, as this approach was found to perform well in comparison with other clustering algorithms for this task.  As the subject columns capture the identifying property of each table, such as the \textit{name} of a \textit{company} or the \textit{title} of a \textit{movie}, clustering by subject column should bring together all the tables representing companies (or perhaps organizations) in one cluster and all the tables representing movies (or perhaps creative works) to provide candidate top-level types. We note that the choice of top-level types is subjective; should \textit{movie} or \textit{creative work} be the top level type? We depend on clustering based on embeddings to make decisions on the granularity of the top-level types, and infer taxonomies to identify finer-grained types. The experiments compare the inferred top-level types of tables to manually annotated tables. The inferred top-level types can be considered to be conceptual types, in that they aim to reflect the concepts represented in the tables.}




\subsection{Identify Attributes}

After identifying the (conceptual) top-level types, we derive the (conceptual) attributes of each type based on the attributes (columns) of the tables in the type's cluster. These table attributes are grouped into clusters, where those within the same cluster are expected to share similar semantics or belong to the same semantic domain. For example, the \textit{location} attribute of an \textit{Organization} table and the \textit{place} attribute of another \textit{Organization} table are likely to represent similar properties of organizations and thus could be consolidated into a single attribute of the top-level type.

Therefore, we apply a clustering algorithm to the embeddings of all the table attributes (columns) of tables of each top-level type, and use the resultant clusters to define its conceptual attributes. 
Note that after this process, each table attribute is mapped to a single conceptual (top-level type) attribute. As in the identification of top-level types, Agglomerative Clustering and Euclidean distance are used, having been shown to provide good performance in experiments.


\vspace{-0.1cm}
\begin{algorithm}[t]
\caption{Dendrogram Pruning}
\label{alg:slice}
\KwData{$dendrogram$; $maxSilhouette$;}
\KwResult{Type Hierarchy}
Initialize $hierarchy$\;
$current_y$ = max(y values in $dendrogram$);\\
\While{$current_y >=0$}{

$clusters$ = getClustersAtHeight($dendrogram$, $current_y$);\\
\If{$getSilhouette(clusters)>(maxSilhouette-\Delta)$}{
 \For {$cluster$ in $clusters$}{
  \If {$cluster$ is not a single table}{
  $tables$ = getTable($cluster$);\\
  $parentCluster$ = findParentCluster($hierarchy$, $cluster$);\\
  \If {$parentCluster \; exists$ }{
   AddEdge($hierarchy$, $parentCluster$, $cluster$);\\}
   \Else{addRoot($hierarchy$, $cluster$)}
  }
 }
}
 $current_y -= \delta$;\\
}
\Return{$hierarchy$}\;
\end{algorithm}

\subsection{Infer Taxonomy}
\label{subsec:infer-hierarchies}

Within each top-level type, tables may reflect different perspectives. For example, a cluster of tables representing  \textit {Organization} might encompass specific subtypes such as \textit{University} and \textit{Company}. In this method, we assume that the subtypes' distinctions are evident through their conceptual attributes, and that a sub-taxonomy can be developed for each top-level type by applying hierarchical clustering to the conceptual attributes that its tables have.
%
%
In particular, for each top-level type $t$, the approach includes:
\begin{enumerate}[leftmargin=*]
\item \textbf{Hierarchical Clustering}: a hierarchical clustering algorithm is applied to the tables of $t$ to construct a dendrogram, where the distance between two tables is the Jaccard similarity of their sets of conceptual attributes. 
\item \textbf{Dendrogram Pruning}: The resulting Dendrogram is sliced (Algorithm~\ref{alg:slice}) at various y-axis levels to generate clusters representing potential subtypes. This approach identifies groups of attributes that co-occur across multiple tables in the cluster. To ensure taxonomy quality, we retain only slices where cluster silhouette scores fall within the range 
$[maxSilhouette - \Delta,\, maxSilhouette]$, where $maxSilhouetteScore$ is the highest silhouette score that can be obtained by slicing the dendrogram and $\Delta$ helps determine the minimum silhouette score for the resulting clusters when slicing the dendrogram.
Each retained cluster is designated as a new subtype in the taxonomy, and the dendrogram's hierarchy is reflected in the taxonomy's structure.
\end{enumerate}

 \begin{figure}[t]
\centering
  \centering
  \includegraphics[width=0.8\linewidth]{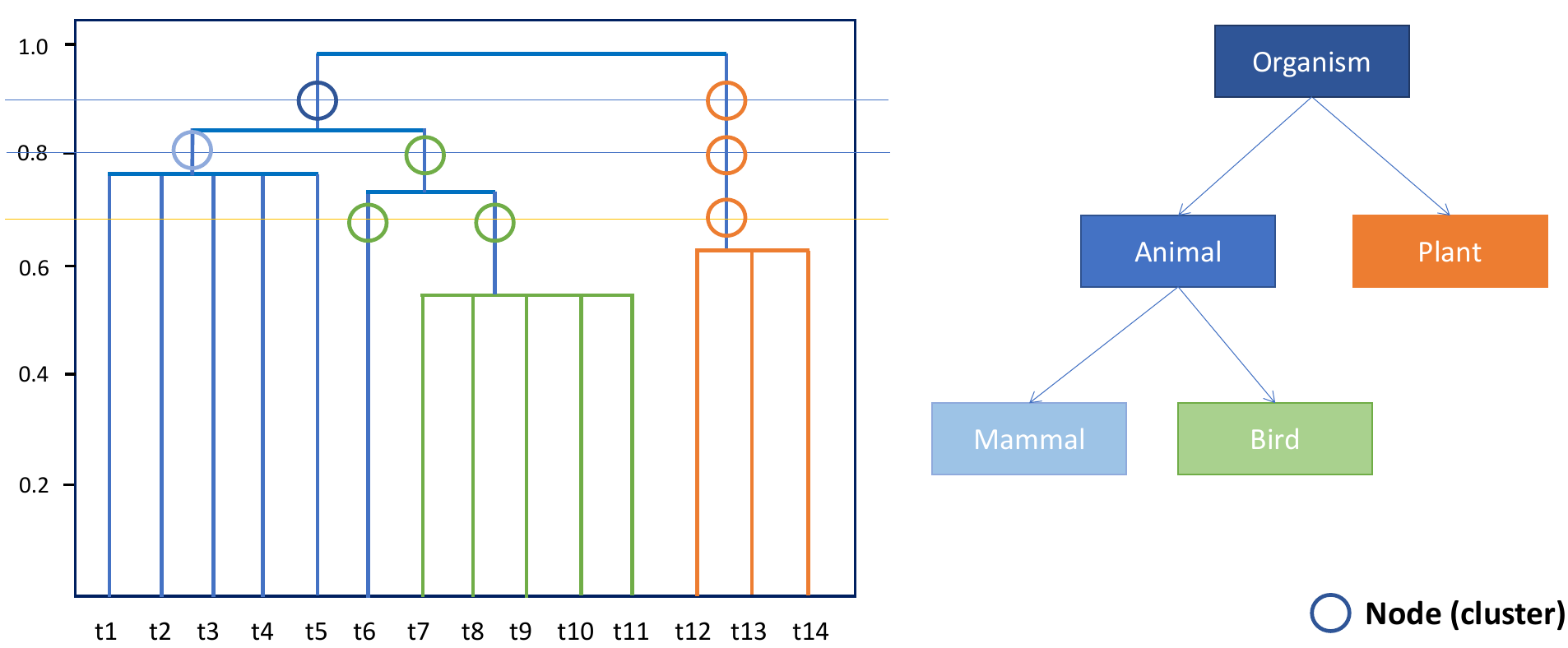}
\caption{{An Example of Dendrogram Prunning from WDC.}}
\label{fig:inheritance-example}
\end{figure}

{Figure \ref{fig:inheritance-example} demonstrates the taxonomy inference process for the top-level type \textit{Organism} in WDC. The dendrogram, generated from clustering 14 tables (\textit{t1} to \textit{t14}), identifies three slices with silhouette scores within $\Delta$ of the highest score.
Each slice is color-coded to match the corresponding types in the GT hierarchy. The slicing algorithm (Algorithm~\ref{alg:slice}) begins at the top of the dendrogram and moves downward, including clusters in the resulting \textit{hierarchy} if their average silhouette scores fall within the predefined range.

In the experiments, $\Delta$ is set to $0.15$ --- a value that has been shown empirically to produce hierarchies that are both at intuitive levels of detail and have good levels of consistency.   Broadly speaking, a higher $\Delta$ leads to a deeper hierarchy.}

\section{Generative Method}
\begin{figure}[t]
    \centering
    \includegraphics[width=1.02\linewidth]{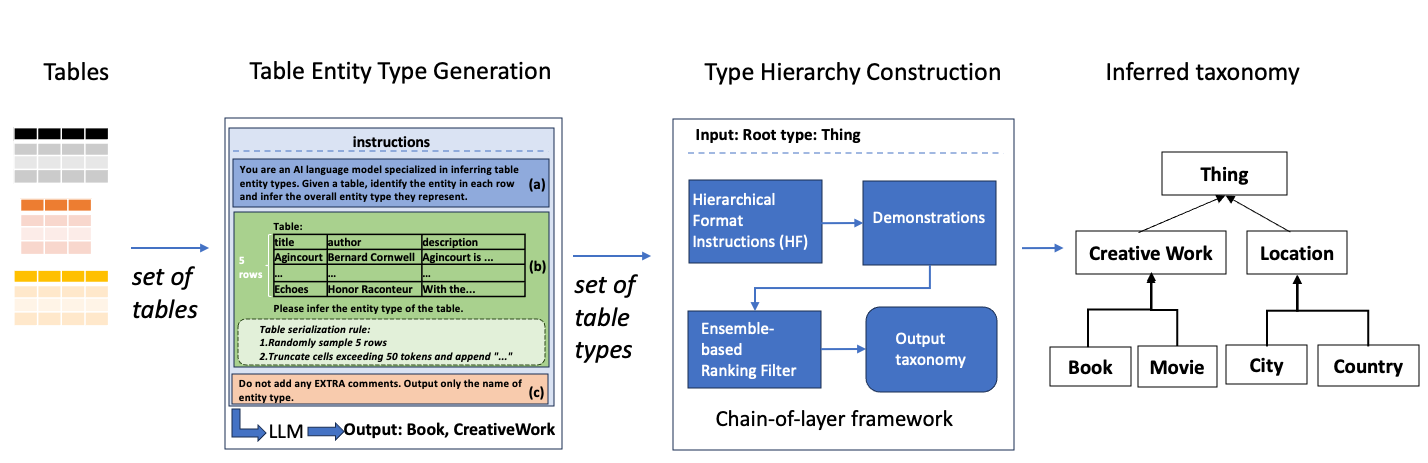}
    \vspace{-0.2cm}
    \caption{The Framework of the Generative Method GeTT}
    \label{fig:gett}
\end{figure}

The framework of GeTT is shown in Figure \ref{fig:gett}. It includes two modules: entity type generation for tables and type hierarchy construction. We will next introduce these two modules with details. 
Note that complete prompts used in both modules can be found in GeTT/prompts.txt in the code and data repository.

\subsection{Table Entity Type Generation}
\label{sec:gett-type}

GeTT uses a generative LLM  to generate potential entity types for each given table, with the structure of the prompt demonstrated in Figure \ref{fig:gett}.
This prompt includes the following parts: (a) a description of the task, (b) a specification of the input including a simple but effective and widely adopted table serialization which uses commas to separate the values \cite{DBLP:conf/naacl/MinHJLCCLQLLW24}, (c) a rule for specifying the output --- solely the names of the entity types.
Current LLMs support a limited context window, which typically encompasses a few thousand tokens. Therefore, a sampling operation needs to be applied to the original table before it is used to construct the prompt. 
Following  \cite{DBLP:conf/vldb/chorus}, we randomly sampled 5 rows from each table. For some columns like ``descriptions'', an individual cell may contain so many tokens that a prompt with one or two rows exceeds the maximum context window size. We thus truncate cells with more than 50 tokens to 50 tokens and append ``...'' to indicate the truncation. {The output of the Table Entity Type Generation step in Figure \ref{fig:gett} is a set of entity type names.
}

\subsection{Type Hierarchy Construction}\label{subsec:CoL}

GeTT first transforms the generated entity types of all the tables into a flat list (denoted as $\mathcal{V}$), where types with the same name are regarded as one type with their associated tables merged, and then feeds $\mathcal{V}$ into an LLM  together with a root type (denoted $v_0$) for constructing a coherent hierarchy. 
{Instead of developing prompts from scratch, we use a state-of-the-art LLM prompt for hierarchy construction named Chain-of-Layer (CoL) \cite{DBLP:conf/cikm/0001BTFLZ024}. There are other methods that construct taxonomies with an LLM (e.g., \cite{DBLP:conf/kbclm/FunkHJL23,DBLP:conf/nips/LoJLJ24}), but their settings are relatively different from ours which has types given and requires no LLM training.}
CoL builds the taxonomy from the top down, starting from the initial layer \(\mathcal{T}^{0}\) (at a top level type resulting from Section~\ref{sec:gett-type}) that is composed of $v_0$ alone, and iteratively adding new layers of types.
At the \(k\)-th iteration, given the current layer \(\mathcal{T}^{k}\), CoL selects the appropriate child types from \(\mathcal{V}\), forming the next layer \(\mathcal{T}^{k+1}\), and removes them from \(\mathcal{V}\). This process continues until \(\mathcal{V}\) becomes empty.
For the technical details of CoL, please refer to \cite{DBLP:conf/cikm/0001BTFLZ024}. Here is a brief introduction to its three main components: 
%
%
%
%
\begin{itemize}[leftmargin=*]
    \item \textbf{Hierarchical Format Instructions:} Each iteration is guided by an instruction that directs the LLM to generate plausible child types of the types in the current layer from the given list \(\mathcal{V}\).
    \item \textbf{Demonstrations:} CoL provides example taxonomies to the LLM.
    They can be taxonomies either annotated by experts or generated by the LLM. Our method GeTT adopts the latter, which is called the zero-shot setting of CoL.
    For each demonstration, the taxonomy is decomposed in a hierarchical order and simulated from top to bottom. After each induction step, the LLM checks if all target entities are included. If not, the taxonomy is further expanded until it encompasses the entire set of entities. 
    \item \textbf{Ensemble-based Ranking Filter:} At each iteration, to mitigate hallucinations, i.e., incorrect or semantically inconsistent parent-child relationships introduced by the LLM, CoL filters out the generated low-quality parent-child relationships by transforming them into sentences with multiple templates and feeding them into a pre-trained mask language model for scoring and ranking. 

\end{itemize}


\section{Evaluation}

\subsection{Experiment Settings}

\subsubsection{Datasets.}
We adopt the following {three} table sets: WDC which includes 602 tables from Web Data Commons ~\cite{springer/ISWC14/WDC} and a Web table set named T2DV2 ~\cite{DBLP:conf/EDBT17/T2DV2}, GDS which includes 660 tables from Google Dataset Search ~\cite{conf/WWW/DanBrickey19} {and OpenData, which includes 10361 tables collected from various Open Data portals worldwide, covering a diverse range of sources beyond those from the UK, US, and Australia \cite{ukopen,usopen,ausopen}.} Each table set is annotated with a ground truth (GT) taxonomy composed of entity types from Schema.org, and each table is also annotated with entity types from Schema.org, {with its most specific entity type and top-level type specified}.
The statistics of the three datasets are shown in Table \ref{tab:benchmark}. Note that lowest-level types are the leaf types in the taxonomy. 

\begin{table}[t]
 \centering
\begin{tabular}{c|c|c|c|c|c|c}
\hline
Dataset & \# Tables & \# Attributes & \# Top-level& \# Lowest-level&\# Entity Types &Depth \\
& \ & \ & \  Types & \  Types &&\\ 
 \hline
WDC       & 602      & 4200      & 7                  & 43           &71&     4    \\ \hline
GDS       & 660      & 15195     & 6                  & 53              &66&      3\\ \hline
OpenData  &10361      &313822    & 6                  & 49             & 62&     3 \\ \hline
\end{tabular}
\vspace{0.1cm}
\caption{Statistics of the Datasets}
\label{tab:benchmark}
\end{table}





\subsubsection{Metrics applied to the Top-level Types.}
Top-level types, as the most fundamental output of schema inference, include important meta information of the table set, and also influence the quality of the constructed taxonomy. 
We regard the top-level type inference as a problem of table clustering, and accordingly calculate the widely used metric
%
$\text{Rand Index (RI)} = \frac{\text{TP} + \text{TN}}{\text{TP} + \text{FP} + \text{FN} + \text{TN}}$,
where TP (resp. TN) is the number of pairs of tables that belong to the same (resp. different) top-level type {in both the output taxonomy $\mathcal{H}_o$ and the GT taxonomy $\mathcal{H}_{gt}$}, and FP (resp. FN)  is the number of pairs of tables that belong to the same (resp. different) top-level type in $\mathcal{H}_o$ but different (resp. same) top-level types in $\mathcal{H}_{gt}$.
We also calculate Purity of each top-level type $t$ in $\mathcal{H}_o$ as the ratio of the tables whose top-level type in $\mathcal{H}_{gt}$
is $m(t)$ (i.e., matched with $t$), among all the tables associated to $t$.
With the Purities of all the top-level types in $\mathcal{H}_o$, we average them as the final Purity of top-level types of $\mathcal{H}_{o}$.

\subsubsection{Metrics applied to the Whole Taxonomy.}
{
We first use some taxonomy statistics, including the number of  types (T\#) and the maximum depth of leaf types, i.e., the number of levels (L\#), for assessment. 
Richer taxonomies have higher T\# and/or L\#.
There have been metrics to assess the correctness of taxonomies but they usually independently assess each edge (i.e., the subsumption of two concepts) ignoring the taxonomy structure, such as the pre-trained language model-based RaTE \cite{langlais2023rate}.
Therefore, we propose a new metric named Tree Consistency Score (TCS) to measure the overall structure consistency between $\mathcal{H}_o$ and $\mathcal{H}_{gt}$, with the basic idea of comparing all the ancestors of two matched concepts in their taxonomies. 
A higher TCS indicates a higher-quality output taxonomy.
It is calculated as follows:
}
\begin{enumerate}[leftmargin=*]
\item For each type $t$ from $\mathcal{H}_o$, we match it with a type that is from $\mathcal{H}_{gt}$ and is the most frequent entity type annotation of the associated tables of $t$. This matched type is denoted as $m(t)$. 
\item We calculate the consistency of each type $t$ from $\mathcal{H}_o$ as 
\begin{equation}
    C_{type}(t) = \frac{|\left\{a \in A(t, \mathcal{H}_o) | m(a) \in   A(m(t), \mathcal{H}_{gt}) \right\}  |}{|A(t, \mathcal{H}_o)|}
\end{equation}
where the function $A(\cdot, \cdot)$ calculates the set of ancestors of a given type in a given taxonomy, and $|\cdot|$ denotes the set cardinality.
\item The TCS score of $\mathcal{H}_o$ is computed as
\begin{equation}
    C_{taxo}(\mathcal{H}_o) = \frac{\sum_{t \in \mathcal{T}_o} C_{type}(t)}{|\mathcal{T}_o|}
\end{equation}
where $\mathcal{T}_o$ is all the types of $\mathcal{H}_o$.
\end{enumerate}



\label{eqn:Purity}

\subsubsection{Embeddings.} 
{For EmTT, we employ the following approaches to create column embeddings, the similarity of which is used to identify top-level types and their attributes, as described in Section \ref{sec:EmTT}:}

\begin{itemize}


\item {SBERT~\cite{DBLP:conf/emnlp/ReimersG19} is a pretrained language model developed for encoding sentences, that has not been fine-tuned for the specific task of column embedding. This is in contrast with the other embedding models used. 
To represent a column as a sentence, we provide a structured format: using $<s>$ as the start token, enclosing the column header in $<header>...</header>$, and appending unique cell values concatenated with spaces.}

\item {\it Starmie} was designed primarily for the Table Union Search problem~\cite{DBLP:journals/pvldb/NargesianZPM18}, which identifies similar table attributes to determine table unionability~\cite{DBLP:journals/pvldb/FanWLZM23}. It applies the SimCLR~\cite{DBLP:conf/icml/ChenK0H20} contrastive learning framework to table attributes, treating table attributes as data items and their subsets as positive variants. Embeddings are generated by fine-tuning RoBERTa~\cite{DBLP:journals/corr/abs-1907-11692} through self-supervised learning to maximize similarity between positive examples and minimize it for negatives.

\item {\it DeepJoin} was developed for discovering joinable attributes, supporting both exact and semantic joins~\cite{DBLP:journals/pvldb/Dong0NEO23}. Positive training examples are derived from attribute pairs with SBERT cosine similarity above $0.9$, while unmatched attributes act as negatives. DeepJoin optimizes embeddings using a contrastive loss function to differentiate between positive and negative examples.


\item {\it Unicorn}~\cite{DBLP:journals/sigmod/FanTLWDJGT24}
{trains an architecture that first encodes data element pairs of multiple matching tasks by a pre-trained language model and an additional Mixture-of-Experts layer, and then predicts the matchings by attaching a classifier.
In our application, Unicorn matches attributes between tables to infer top-level types and conceptual attributes.}
Matched pairs are represented as edges in a graph, where nodes correspond to attributes. 
Attributes are grouped into clusters as connected subgraphs, with each cluster treated as a top-level type's attribute.

\item {{\it SwAV} ~\cite{DBLP:conf/nips/CaronMMGBJ20}, is a contrastive learning technique originally developed for image embedding and clustering~\cite{DBLP:conf/nips/CaronMMGBJ20}; here we apply it for the first time to tabular data. Unlike traditional contrastive learning approaches such as SimCLR~\cite{DBLP:conf/icml/ChenK0H20}, which rely heavily on both positive and negative samples and require pairwise feature comparisons within large batches, SwAV reduces the need for explicit negative pairs and large batch sizes by learning consistent cluster assignments across multiple augmentations (views) of the same data item. Here, we use SwAV to fine tune SBERT embeddings for column comparison.}
\end{itemize}

\subsubsection{Other Features of Experiment Setup.} 
EmTT and GeTT are implemented using PyTorch, the Hugging Face Transformers, and the Ollama library.
{Details of the fine-tuning in EmTT with Starmie, DeepJoin and SwAV are presented in the Appendix \ref{sec:existing-lms}}.
{We evaluate GeTT using two closed-source LLMs (GPT-3.5 and GPT-4) and three open-source LLMs, namely deepseek-R1 and two different-sized variants of Qwen2.5\footnote{Llama 3 was also tested but inconsistent output representations made it difficult to incorporate in the Chain-of-Layer workflow.}, as they are widely adopted and often achieve promising results.}

We run each configuration 5 times, reporting the averages as well as the standard deviations.
%
To run EmTT and GeTT with Qwen2.5 (14B/32B), which rely on local GPU, we use an NVIDIA A100 (80GB)  with 2 x 24-core AMD Epyc 7413 2.65GHz processors and 512GB RAM. For GeTT with GPT-3.5/4, we use a workstation with a 14-core Intel Xeon E5-2680 v4 processor with 128GB RAM.

\subsubsection{Availability.} All the datasets and code can be accessed from:  \newline\url{https://github.com/PierreWoL/TwoMethods}.
 

\subsection{Result Analysis}
The experimental results are reported in Tables \ref{tab:overallResult}, \ref{tab:std} and \ref{tab:time_comparison}. We will next analyse the quality of the top-level types and the overall taxonomies, and compare EmTT and GeTT w.r.t. efficiency and stability.

 \begin{table}[t]
\centering
 \resizebox{1\linewidth}{!}{
\begin{tabular}{l|ccccc|ccccc|ccccc}
\hline
\textbf{Method} & \multicolumn{5}{c|}{\textbf{WDC}} & \multicolumn{5}{c|}{\textbf{GDS}}  & \multicolumn{5}{c}{\textbf{OpenData}}\\
\cmidrule(lr){2-6} \cmidrule(lr){7-11}\cmidrule(lr){12-16}
& \textbf{TCS}  & \textbf{T\#}  & \textbf{L\#}     & \textbf{RI} & \textbf{Purity} & 
\textbf{TCS}&  \textbf{T\#}&  \textbf{L\#} & \textbf{RI} & \textbf{Purity} &
\textbf{TCS}&  \textbf{T\#}&  \textbf{L\#} & \textbf{RI} & \textbf{Purity}\\
\hline
EmTT (SBERT)
&  \textbf{1.000}  &158&  \underline{5}      & 0.770 & 0.814 
&\textbf{ 0.929} &231&   5    & 0.792  & 0.674&\textbf{0.845} & 936&\underline{7}& 0.889& 0.884\\ 
EmTT (Starmie) & \textbf{1.000} & 117& 2 & 0.771& 0.754
&\underline{0.895} &  134& 3 & 0.813& 0.726
&0.813 & 743& 5 & 0.841& 0.772\\
EmTT (DeepJoin) & 0.836& 162 & 4 & 0.845 & 0.842
& 0.848& 214& 5 & \textbf{0.896} & 0.857
&\underline{0.821}& 897& \underline{7} & 0.879 & 0.878\\
EmTT (Unicorn) & -&-& - &0.734& 0.890  
& -&-&-& 0.709& \textbf{0.903} 
& -&-& -& -& - \\
EmTT (SwAV) & \underline{0.856}&  213 & \textbf{6}& 0.865 &0.870
& 0.800&276&   \textbf{7} & \underline{0.885}&  0.856
&0.783 &1228 & \textbf{10}& 0.881&0.851 \\
\hline
GeTT (GPT-3.5) &0.490 &354 &  4  & 0.516 & 0.747
& 0.437 & 293  &  5   & 0.852  & 0.526 
&0.457& 1034  &  5  &0.879  & 0.914 \\
GeTT (GPT-4)&{0.816}  &\textbf{358}&  \underline{5}  &\underline{ 0.976}& \textbf{0.961}  
&0.425&296&  5  & 0.760&\underline{0.900} 
&0.516&1162&  6  &\textbf{ 0.912}&\underline{0.906}  \\
GeTT (DeepSeek-R1)&0.813&348 &5 & \textbf{0.980} & \underline{0.957}&
0.596&\textbf{ 302}&  5  & 0.819&0.787 &
0.532& 1216& 6  & \underline{0.907 }& 0.898 \\
GeTT (Qwen2.5-14b)
& 0.602 &352& 4 & 0.727 & 0.738
 & 0.562&294&  4 & 0.784& 0.610 
 & 0.453&\textbf{1344}& \underline{7}& 0.836& 0.797\\
GeTT (Qwen2.5-32b)
&  0.681 &\underline{356}&   \underline{5}  & 0.814  & 0.930 &
0.493&\underline{295}&  \underline{6}  & 0.751  & 0.673 &
0.545&\underline{1258}& 6 &0.809 & 0.782\\
\hline
\end{tabular}
}
\vspace{0.1cm}
\caption{Results of the baselines, EmTT and GeTT over all the metrics, with the best results in bold and the second best underlined.}
\label{tab:overallResult}

\end{table}

\subsubsection{Quality of the Top-level Types} {In EmTT, among the embedding methods, it can be seen from Table~\ref{tab:overallResult} that DeepJoin and SwAV provide the most consistently strong performance for RI and Purity.}
{SwAV fine tunes SBERT for column similarity, so its advantage w.r.t. SBERT stems from its contrastive learning mechanism}, which pulls attributes with mutual information closer together in the embedding space, resulting in small, dense and high-purity clusters. DeepJoin benefits from the use of (automatically identified) positive attribute examples during training, which helps the model embed similar attributes closer together. In contrast, Starmie, which is based on SimCLR, struggles to separate attributes that share overlapping values but have different meanings. For instance, \textit{music album} names such as \textit{The Mozart Album} and \textit{Mozart Momentum: 1785} may be confused with \textit{event} names like \textit{Mozart Gala} or \textit{Mozart's Violin Concerto No. 5}. This issue arises because Starmie’s loss function maximizes the distance between negative samples within each batch but does not account for negatives across batches. As a result, Starmie sometimes fails to distinguish semantically distinct but plausible attributes, leading to fewer, larger clusters that mix attribute types and yield lower Rand Index scores compared to EmTT. 

{GeTT's performance varies significantly on different datasets, depending on the adopted LLMs. On the WDC dataset, GeTT (DeepSeek-R1) achieves a RI of 0.980 and Purity of 0.957, while GeTT (GPT-4) attains 0.976 and 0.961; these values surpass EmTT (SwAV) by approximately 10\%. By contrast, GeTT with Qwen2.5-14b produces more modest RI and Purity scores of 0.727 and 0.738, which fall below most EmTT embedding models. 
On the GDS dataset, GeTT (DeepSeek-R1) trails EmTT (DeepJoin) by roughly 8\% in RI, whereas GeTT (GPT-4) achieves a Purity of 0.900, comparable to EmTT (Unicorn). Nevertheless, GeTT with Qwen2.5-14b/32b  remains slightly below EmTT’s low-performing variants on both RI and Purity.
On OpenData, GeTT (GPT-4) and GeTT (DeepSeek-R1) achieve the highest Rand Index (0.912 and 0.907), whereas GeTT (GPT-3.5) and GeTT (GPT-4) yield the top Purity (0.914 and 0.906).  

Overall, GeTT produces the best Rand Index and Purity with DeepSeek-R1 and GPT-family models for WDC and OpenData, but EmTT (DeepJoin) and EmTT (SwAV) are competitive in GDS.}


\subsubsection{Quality of the Taxonomies}
{In Table \ref{tab:overallResult} all of TCS, T\# and L\# report features of the generated taxonomies. 

In EmTT, it is noteworthy that different methods produce different numbers of types (T\#) and depths of hierarchy (L\#). For example, on all three datasets, EmTT (SwAV) generates more types and layers than with the other embeddings. This means that EmTT (SwAV) often produces richer hierarchies than Starmie or DeepJoin}.  Unicorn fails to complete the type taxonomy construction due to scalability limitations.

Increased taxonomy complexity from EmTT (SwAV) comes at a cost in terms of the tree consistency score. EmTT (SwAV) creates more small column clusters, where each cluster corresponds to a specific top-level type attribute. 
This leads to a more complex hierarchy driven by shared attributes. This richer hierarchy, however, provides more opportunities for mismatches with the GT taxonomy, leading to a lower tree consistency score. In GDS, for instance, \textit{School} is sometimes misclassified as a subtype of \textit{LocalBusiness} because both types share similar address-related attributes. Notably, when no hierarchy is constructed under a top-level type, the tree consistency score defaults to \textit{1}, indicating perfect consistency. 





\begin{figure}[t] 
  \centering 
  \subfigure[Taxonomy with Abstract Types]{ 
    \label{fig:GeTTe1}
    \includegraphics[width=2.15in]{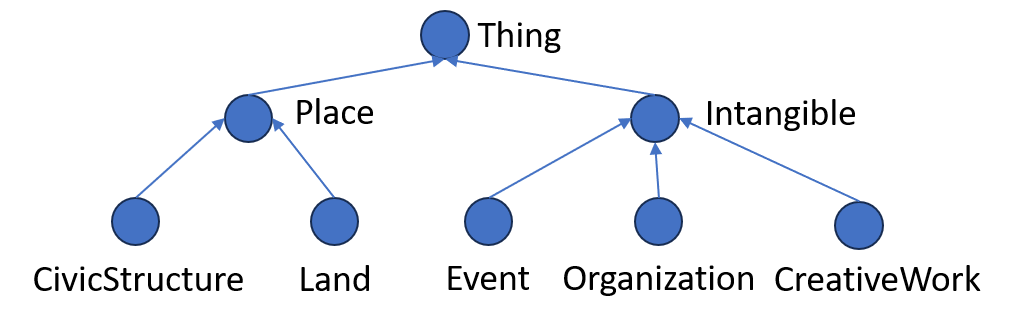} 
  } 
  \subfigure[Taxonomy with Specific Types]{ 
    \label{fig:GeTTe2}
    \includegraphics[width=2.3in]{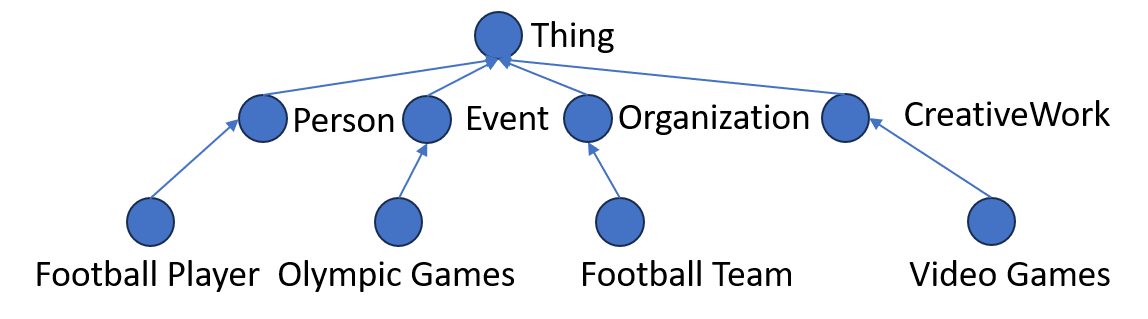} 
  } 
  \vspace{-0.2cm}
  \caption{Examples of the top-3 levels of the inferred taxonomies by two separated runs, using GeTT (GPT-4) on the GDS dataset. 
  } 
  \label{fig:GeTTe} 
\end{figure}

{Compared to EmTT with all embedding methods, GeTT infers more types but with reduced overall tree consistency. 
GeTT generates entity types directly from tables, resulting in more types than EmTT in most cases. Specifically, with four different LLMs, GeTT in average generates 353.6 types on WDC, 296 on GDS and 1202.8 on OpenData --- 66.7\%, 7.8\% and 2\% more than EmTT (SwAV), respectively.}
For example, while EmTT (SwAV) clusters three tables into a single type category representing \textit{Healthcare Facility}, GeTT infers more granular types like \textit{Healthcare Facility}, \textit{Hospital}, and \textit{Medical Clinic}. 
The taxonomy depth of GeTT with all language models is comparable to that of EmTT (SBERT), aligning more closely with the ground truth layers, though slightly shallower. The increased semantic granularity of inferred types can lead to inconsistencies when the types become too specific or overly abstract, resulting in instability over taxonomy consistency. As a result, GeTT underperforms in consistency compared to EmTT with different embedding models. 

Figure \ref{fig:GeTTe} illustrates that performance instability arises when the LLM infers overly abstract types. 
In Figure \ref{fig:GeTTe1}, the LLM infers an abstract type \textit{Intangible}, which serves as a top-level type subsuming multiple semantically distinct subtypes.
This abstract type groups plausible but unrelated types together, causing difficulties in distinguishing between levels in the taxonomy. In contrast, Figure \ref{fig:GeTTe2} demonstrates an example where the LLM infers more specific top-level types that better capture the semantics of the underlying table. As a result, the generated taxonomy is clearer and more aligned with the expected GT taxonomy. 
This highlights a limitation of the GeTT method: it relies heavily on organizing types inferred by the LLM without adequately considering the overall semantic content of the tables. When the types are overly abstract, the taxonomy is likely to lack support from actual table content.
Consequently, the quality of the generated taxonomy depends on the accuracy and granularity of the inferred table types.

\begin{table}[t]
\centering
 \resizebox{1\linewidth}{!}{
\begin{tabular}{l|ccc|ccc|ccc}
\hline
\multirow{2}{*}{\textbf{Method}} & \multicolumn{3}{c|}{\textbf{WDC}} & \multicolumn{3}{c}{\textbf{GDS}} & \multicolumn{3}{c}{\textbf{OpenData}}\\
\cline{2-10}
& RI-STD & Purity-STD & TCS-STD & RI-STD & Purity-STD & TCS-STD& RI-STD & Purity-STD & TCS-STD \\
\hline
EmTT (SBERT) & 0  &0 & 0 
& 0  & 0 & 0
& 0  & 0 & 0\\  
EmTT (SwAV) &0.013& 0.012 & 0.021 & 
0.011 & 0.013 & 0.017
& 0.009&  0.014 & 0.021\\\hline

GeTT (GPT-3.5)  & 0.091  & 0.090 & 0.111& 0.134 & 0.129 & 0.138&
0.112&  0.067 &0.146\\
GeTT (GPT-4)  & 0.015 & 0.040 & 0.243
 & 0.245 & 0.163 &0.102 &
0.058 & 0.073 & 0.103 \\
 GeTT (DeepSeek-R1)  & 0.087 &0.033& 0.089
 & 0.104&  0.094  &0.126 &
0.067 & 0.093 &0.117 \\
GeTT (Qwen2.5-14b) & 0.164 & 0.265 & 0.206 & 0.178 & 0.349  & 0.288&
0.091& 0.104 &  0.121\\
GeTT (Qwen2.5-32b)& 0.089 & 0.056 
& 0.177 & 0.162 & 0.174 & 0.233 &
  0.059&  0.060 & 0.183 \\
\hline
\end{tabular}}
\vspace{0.1cm}
\caption{Results of standard deviation (STD) of EmTT and GeTT}
\label{tab:std}
\end{table}

 \begin{table}[t!]
\centering
 \resizebox{1\linewidth}{!}{
\begin{tabular}{lccc}
\toprule
\textbf{Method} & \multicolumn{3}{c}{\textbf{Total Time}} \\
\cmidrule(lr){2-4}
                & \textbf{WDC} & \textbf{GDS} & {\textbf{OpenData}}\\ \midrule
EmTT (SBERT)& 4.59 min & 7.61 min  &594.44 mins\\
EmTT (Unicorn) & 45.21 min  & 54.84 min &  - \\ 
EmTT (Starmie) & 202.86 mins (train: 200 mins)  & 207.74  mins (train: 201 mins) & 1532.75 min (train: 927 min)\\ 
EmTT (DeepJoin) & 339.41 mins (train: 334  mins)  & 415.79  mins (train: 408 mins) & 2241.26 min (train:1647 min)\\ 

EmTT (SwAV) & 239.52 mins (train: 234 mins)  & 277.81 mins (train: 270 mins) & 2156.75 min (train: 1591.65 min)\\ \midrule
GeTT (GPT-3.5)   & 11.08 mins  & 13.18 mins & 164.72 min  \\
GeTT (GPT-4)     & 17.26 mins  & 20.62 mins &135.13 min \\
GeTT (DeepSeek-R1)     & 65.28 mins & 70.83 mins  & 343.97 min \\
GeTT (Qwen2.5-14b) & 12.58 mins & 14.69 mins   & 95.64 min \\
GeTT (Qwen2.5-32b) & 21.27 mins & 26.51 mins &  109.84 min \\  \bottomrule
\end{tabular}}
\vspace{0.1cm}
\caption{Overall Running Times for EmTT and GeTT. }
\label{tab:time_comparison}
\end{table}

\subsubsection{Efficiency and Stability}
We compare the stability of GeTT and EmTT across metrics relating to top-level types (RI and Purity) and the taxonomy (TCS) in Table \ref{tab:std}. For TCS, GeTT exhibits significant variability, with standard deviations ranging from {0.089} to 0.288 across {three} datasets, far exceeding those of EmTT. Similarly, for top-level type inference, 
GeTT again shows higher variability with the standard deviation of the RI ranging from {0.104 to 0.349 on GDS, 0.058 to 0.112 on OpenData}.
This indicates that GeTT is short of robustness in the generation of taxonomies.
Such instability without repeatable performance in LLM prompting-based methods is reported in other studies \cite{DBLP:journals/corr/abs-2408-14595,DBLP:journals/corr/abs-2408-04667}. 
{Meanwhile, as the inference steps in EmTT are deterministic, 
EmTT (SBERT)} has standard deviations of 0. Even when EmTT involves fine-tuning large language models multiple times and then encoding, as in EmTT (SwAV), the resulting performance fluctuations remain very small, with standard deviations only around 0.01–0.02.

As for the efficiency, the overall running times of GeTT are much less than those of EmTT, where fine-tuning is required, as shown in Table \ref{tab:time_comparison}. 
{The EmTT variations, except those using SBERT and Unicorn, require a fine-tuning phase that costs about 3.5–4.5 hours on smaller datasets like WDC and GDS, and 16–25 hours on the large-scale OpenData. 
For example, when running on the same computational resource, GeTT (Qwen2.5-14b) takes only around 5\% of the overall time of EmTT across the three datasets. Even when a larger Qwen2.5 model is applied locally or an online closed-source LLM like GPT-4 is used, the running time of GeTT remains significantly lower than that of EmTT where fine-tuning is required. Furthermore, for large datasets with a high number of attributes—such as OpenData, which contains 310K attributes—EmTT still exhibits noticeably longer clustering times, even when employing a pre-trained LM like SBERT, requiring 1.5–5 times more time than GeTT variations. Meanwhile, the Unicorn variant of EmTT incurs a high computational cost due to its reliance on pairwise attribute matching, making it unable to complete all tests on OpenData.
}

\section{Related Work}

\noindent\textbf{Schema Inference}. A relevant line of work to taxonomy inference for tabular data is schema inference.  Schema inference techniques have primarily focused on semi-structured data formats such as XML, JSON, and RDF. These techniques produce output schemas that range from high-level integration schemas~\cite{DBLP:journals/pvldb/KhatiwadaSGM22} and concise summaries~\cite{DBLP:conf/vldb/YuJ06a,DBLP:journals/pvldb/YangPS09} to full disjunctions capturing structural and semantic patterns~\cite{DBLP:journals/pvldb/KhatiwadaSGM22}. 
For XML and JSON, schema inference often relies on element names and structural patterns identified through graph-based partitioning~\cite{DBLP:conf/edbt/BarretMU24}. In contrast, RDF schemas leverage type annotations to handle heterogeneous and inconsistent data sources~\cite{DBLP:journals/tlsdkcs/ChristodoulouPF15,DBLP:conf/er/Kellou-MenouerK15}, while some approaches employ similarity metrics to merge distinct schemas~\cite{DBLP:journals/semweb/FloresRNGRJD24}.
Our work addresses a gap in schema inference: inferring a conceptual taxonomy specifically tailored to heterogeneous tabular datasets. 
Unlike existing methods that rely heavily on consistent naming or type annotations, our approach EmTT uses LLMs to embed column-level semantics, which accommodates inconsistent terminologies and encodes nuanced contextual clues, thereby enabling similarity-based type inference and the construction of a taxonomy, while our another approach GeTT directly generates table semantics of entity types in an end-to-end way using generative LLMs.

\noindent \textbf{Column Semantics}. Language model-based encoding is increasingly leveraged to capture column semantics across various data integration tasks beyond schema inference.
For instance, Unicorn~\cite{DBLP:journals/sigmod/FanTLWDJGT24} learns embedding-based classifiers for schema matching and column type annotation; DeepJoin~\cite{DBLP:conf/icde/DongT0O21} fine-tunes a pretrained language model to identify joinable columns;
and Starmie~\cite{DBLP:journals/pvldb/FanWLZM23} leverages column-level semantics for data search and table compatibility.
We have included these methods, {along with the first use of SwAV~\cite{DBLP:conf/nips/CaronMMGBJ20} with tabular data, in our experiments, allowing comparison of their effectiveness on a new problem}.

\noindent\textbf{LLM-based Taxonomy Construction}. Recently, methods like Chain-of-Layer (CoL) \cite{DBLP:conf/cikm/0001BTFLZ024} and TaxonomyGPT \cite{DBLP:conf/MODELS23/TaxonomyGPT} have exploited the in-context learning capabilities of LLMs to construct taxonomies. 
CoL, in particular, employs an iterative prompting strategy and a filter module to reduce LLM's hallucination, improving the reliability of the inferred taxonomy. 
Our approach GeTT uses an LLM to infer each table's entity types and leverages CoL to assemble these types into a hierarchical structure. It is worth mentioning that while some prior methods infer entity types from textual contexts \cite{DBLP:conf/acl/emnlp21/FETlabel,DBLP:conf/acl/COLING12/HYENA}, including zero-shot approaches \cite{DBLP:conf/acl/COLING16/labelEmbedFET,DBLP:conf/acl/NAACL19/fetDesc,DBLP:conf/AAAI18/OTyper,DBLP:conf/acl/coling20/MZET}, they are not designed for tabular data, which include semantics of not only text and data values but also structures with rows representing entities and columns representing their attributes.
Among existing methods, Chorus \cite{DBLP:conf/vldb/chorus} is the closest to ours. However, it uses predefined ontology classes, while our approaches stand out as open-domain solutions that directly infer table taxonomies using LLMs, without relying on predefined semantics, enabling more generalizable taxonomy construction across diverse structured data sources.

\section{Conclusion and Future Work}
In this study, we propose two LLM-based methods for taxonomy inference with a given set of tables: EmTT which is based on similarities and hierarchical clustering over column embeddings achieved by fine-tuning encoding LLMs; and GeTT which generates entity types for the tables and organizes the types by prompting decoding LLMs. {Our empirical evaluations show that both EmTT and GeTT can infer appropriate top-level types and taxonomies that show good consistency with manually annotated tables. However, there are interesting differences within and between the methods. In EmTT, the embedding method used has a significant impact on the richness of the taxonomy. In GeTT, both the language model used and the dataset have a significant bearing on result quality. Overall, while GeTT-based proposals using GPT-4 and DeepSeek-R1 have tended to give rise to the highest scores for top-level type inference, EmTT (SwAV) and EmTT (DeepJoin) have provided more dependable performance for taxonomy inference.
GeTT relies on no training with higher efficiency than fine-tuned EmTT methods, but it is relatively unstable. Furthermore, fine-tuned EmTT solutions can be used for different tasks and datasets. } In the future, we will explore more robust frameworks with an ensemble of effective prompts and multiple LLMs, and consider instruction tuning decoder-alone LLMs for not only higher-quality output taxonomies but also higher stability.
We will also explore other related tasks in schema inference including learning type level relationships.


\bibliographystyle{plain}

\bibliography{vada}
\newpage
\appendix

\section{Fine-tuning Configurations in EmTT}
\label{sec:existing-lms} 
\subsection{SwAV and Starmie Configurations}
To investigate the optimal parameter configurations of the column encoder of EmTT, we used the WDC dataset and evaluated the performance of the column encoder model based on the top-level type inference results. 

SwAV and Starmie are both based on constrastive learning, which are self-supervised, allowing for the use of the same datasets for both training and testing, and this is the approach taken in the experiments. For training, the following parameters are used with the AdamW optimization algorithm: Batch Size: 64; Epochs: 100 (SwAV)/64 (Starmie); Contrastive loss: decay to $1e^{-6}$; Learning rate: $5e^{-5}$;

For data augmentation in Starmie and SwAV, we set the number of views ($t$) to 2 and sampled cells in proportion to their TFIDF value as the data augmentation operator ($OP$).  In SwAV, we adopted a $Serial$ strategy that includes both the header and the cells, serializing each column into a string. The choices for $OP$ and $Serial$ strategy were determined by evaluating each parameter individually while keeping all other parameters fixed. 

To determine the optimal sampling fractions $(p_1, p_2)$ for generating the two views of a column during fine-tuning, we started with $(0.5, 0.5)$ to align with Starmie's initial setting. We then empirically tested additional pairs, including $(0.5, 0.3)$, $(1.0, 0.5)$, $(0.8, 0.4)$, $(0.6, 0.3)$, $(0.5, 0.25)$ and $(0.3, 0.3)$, and found that \textbf{$(p_1, p_2) = (0.5, 0.3)$} achieved the best performance. Detailed experimental results are provided in ``README.md'' in our data and code repository.

Regarding the parameters for online clustering and the loss function in SwAV, 
{we set the number of prototypes ($K$) to 500 for training only subject attributes and 3000 for all attributes on WDC and GDS. For OpenData, we use 8000 for subject attributes and 30000 for all attributes.}
The recommended practice from original SwAV~\cite{DBLP:conf/nips/CaronMMGBJ20} framework is to initialise $K$ around an order of magnitude larger than the expected number of clusters. 
The dimension of the projection head in the model (\(d\)) is set to 768, implemented as a sequence of fully connected layers.  We use a temperature coefficient ($\tau$) of 0.07, consistent with Starmie's settings \cite{DBLP:journals/pvldb/FanWLZM23}. The parameter $\epsilon$ in the Sinkhorn algorithm, which controls the smoothness of the assignment process during online clustering, was set to 0.03, as suggested by the SwAV framework.

\subsection{DeepJoin Configurations}
{For DeepJoin, we follow the original configuration from its paper, using a batch size of 32, a learning rate of $2e^{-5}$, and a weight decay of 0.01. The model is trained for 25 epochs with SBERT (all-mpnet-base-v2) as the backbone. Additionally, we set the shuffle rate to 0.3 and adopt the “colname-stat-col” format for column-to-text processing.}

\end{document}